\documentclass[10pt,journal,compsoc]{IEEEtran}
\ifCLASSINFOpdf
\else
\fi
\usepackage[space,nocompress]{cite}
\usepackage{graphicx,caption,amssymb,amsmath,array}
\usepackage{color,float}
\usepackage{subfigure,epsfig}
\usepackage{ragged2e}
\usepackage{enumitem}
\usepackage{footnote}

\usepackage{algorithm}
\usepackage{algorithmic}
\usepackage{multirow}
\usepackage{float}
\usepackage{url}
\usepackage{enumitem}
\usepackage{array}
\usepackage{booktabs}
\usepackage{bm}
\usepackage{color}

\setlist[itemize]{leftmargin=*}
\setlist[enumerate]{leftmargin=*}
\begin{document}
\title{Towards Micro-video Thumbnail Selection via a Multi-label Visual-semantic Embedding Model}


% \author{\IEEEauthorblockN{Yinwei\IEEEauthorrefmark{1},
% Xiang Wang\IEEEauthorrefmark{1},
% Xiangnan He\IEEEauthorrefmark{2},~\IEEEmembership{Member,~IEEE}, \\
% Liqiang Nie\IEEEauthorrefmark{3},~\IEEEmembership{Member,~IEEE}, 
% and Tat-Seng Chua\IEEEauthorrefmark{1},~\IEEEmembership{Member,~IEEE}}
% \IEEEauthorblockA{\IEEEauthorrefmark{1}School of Computing, National University of Singapore, Singapore.}
% \IEEEauthorblockA{\IEEEauthorrefmark{2}, University of Science and Technology of China, Hefei, China}
% \IEEEauthorblockA{\IEEEauthorrefmark{3}College of Computer Science and Technology, Shandong University, Qingdao, Shandong, China.}
% \thanks{Xiang Wang and Liqiang Nie are the corresponding authors.}}
\author{Bo~Liu~\IEEEmembership{, Member, IEEE}
}

\markboth{Journal of \LaTeX\ Class Files,~Vol.~14, No.~8, August~2015}%
{Shell \MakeLowercase{\textit{et al.}}: Bare Demo of IEEEtran.cls for IEEE Transactions on Magnetics Journals}



% for Transactions on Magnetics papers, we must declare the abstract and
% index terms PRIOR to the title within the \IEEEtitleabstractindextext
% IEEEtran command as these need to go into the title area created by
\IEEEtitleabstractindextext{%
\justifying  
\begin{abstract}
The thumbnail, as the first sight of a micro-video, plays pivotal roles in attracting users to click and watch. While in the real scenario, the more the thumbnails satisfy the users, the more likely the micro-videos will be clicked.
In this paper, we aim to select the thumbnail of a given micro-video that meets most users’ interests. Towards this end, we present a multi-label visual-semantic embedding model to estimate the similarity between the pair of each frame and the popular topics that users are interested in. In this model, the visual and textual information is embedded into a shared semantic space, whereby the similarity can be measured directly, even the unseen words. Moreover, to compare the frame to all words from the popular topics, we devise an attention embedding space associated with the semantic-attention projection. With the help of these two embedding spaces, the popularity score of a frame, which is defined by the sum of similarity scores over the corresponding visual information and popular topic pairs, is achieved. Ultimately, we fuse the visual representation score and the popularity score of each frame to select the attractive thumbnail for the given micro-video. Extensive experiments conducted on a real-world dataset have well-verified that our model significantly outperforms several state-of-the-art baselines.
\end{abstract}

}

\maketitle

\IEEEdisplaynontitleabstractindextext

% \fi
\IEEEpeerreviewmaketitle

\section{Introduction}
Micro-video~\cite{MMGCN, GRCN, MGAT}, as a new media type, allows users to record their daily life within a few seconds and share over social media platforms (e.g., Instagram\footnote{https://www.instagram.com/}, and Tiktok\footnote{https://www.tiktok.com/}). 
As the most representative snapshot, the thumbnail summaries the posts and provides the first impression to the observation~\cite{1}. To automatically obtain the superior thumbnail, the prior studies roughly fall into two groups: monomodal-based and multimodal-based methods. In particular, the monomodal-based methods~\cite{3,4,5} merely use the visual information to represent the posts. While, the later ones always incorporate the visual content with the multi-modal information, like titles and descriptions, in order to improve the posts' representation. 
\begin{figure}
	\centering
	  \includegraphics[width=0.47\textwidth]{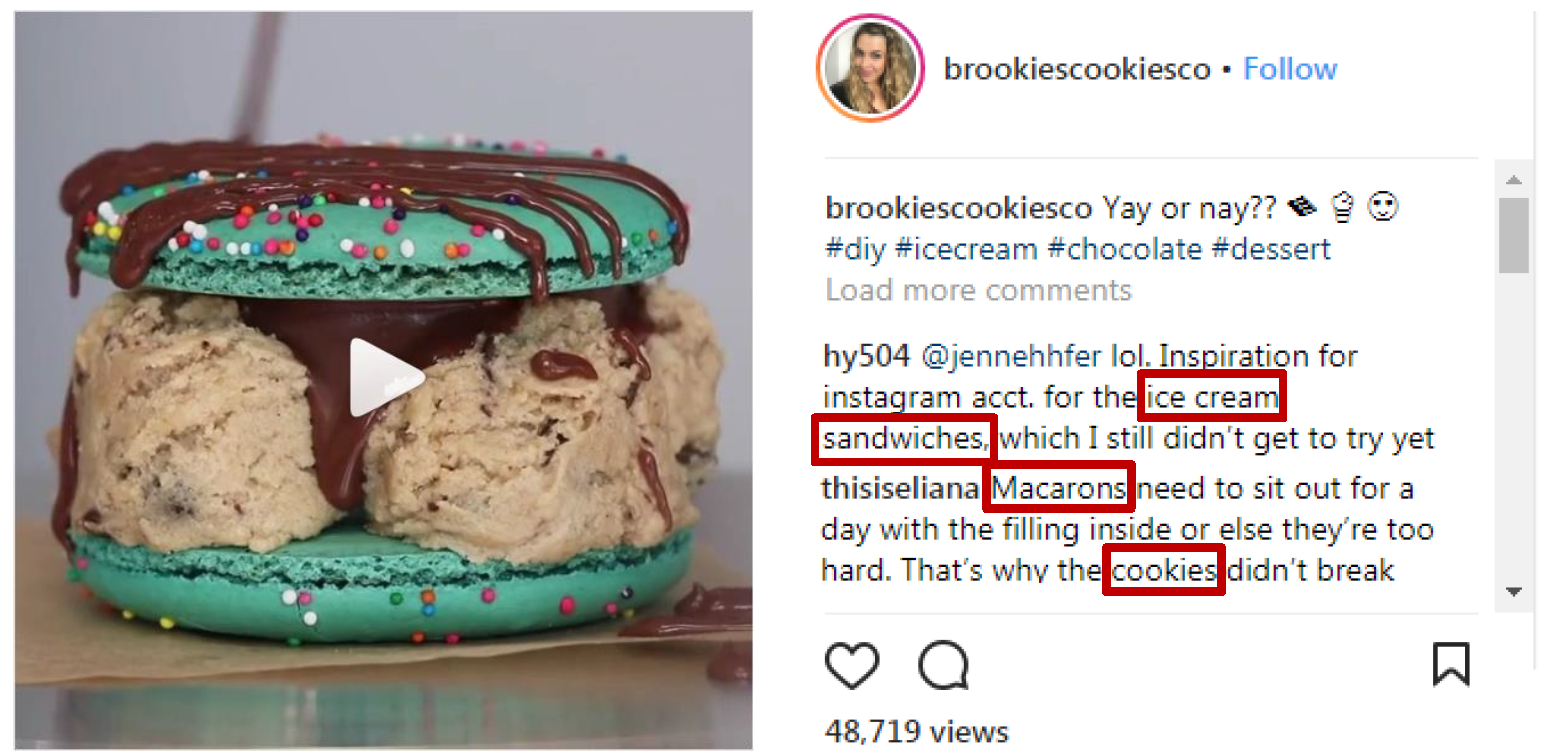}
	  \vspace{-1mm}
	  \caption{Demonstration of user interests reflected by the comments. The red boxes are the entity words.}
	\label{fig_1}	  
	\vspace{-5mm}
\end{figure}

However, we argue that they ignore the comments from the observers for the posts, which reveals their intents towards the posts. Specially, as shown in Figure 1, the words “\textit{ice cream}”, “\textit{Macarons}” and “\textit{cookies}”, reflect the observers' interests to the dessert. Therefore, we propose to discover the observations' interests distribution from their comments and select the frame meeting the majority of observers' intents as the thumbnail of micro-video. 

However, it is non-trivial due to the following challenges: 1) There is no such a dataset can be used to train and evaluate the model. Considering that the observers' intents are dynamic and the manual annotation is high-cost, it is hard to label the extracted frames of the micro-videos for training. And 3) in the cases of the unseen words, emerging in the popular topic list after the training phase, the standard cross-modal similarity calculation approaches cannot be employed.
Especially, one frame may contain various concepts or topics, and the related information is often mixed, which probably confuses the similarity calculation.
\begin{figure*}
	\centering
	  \includegraphics[width=1\textwidth]{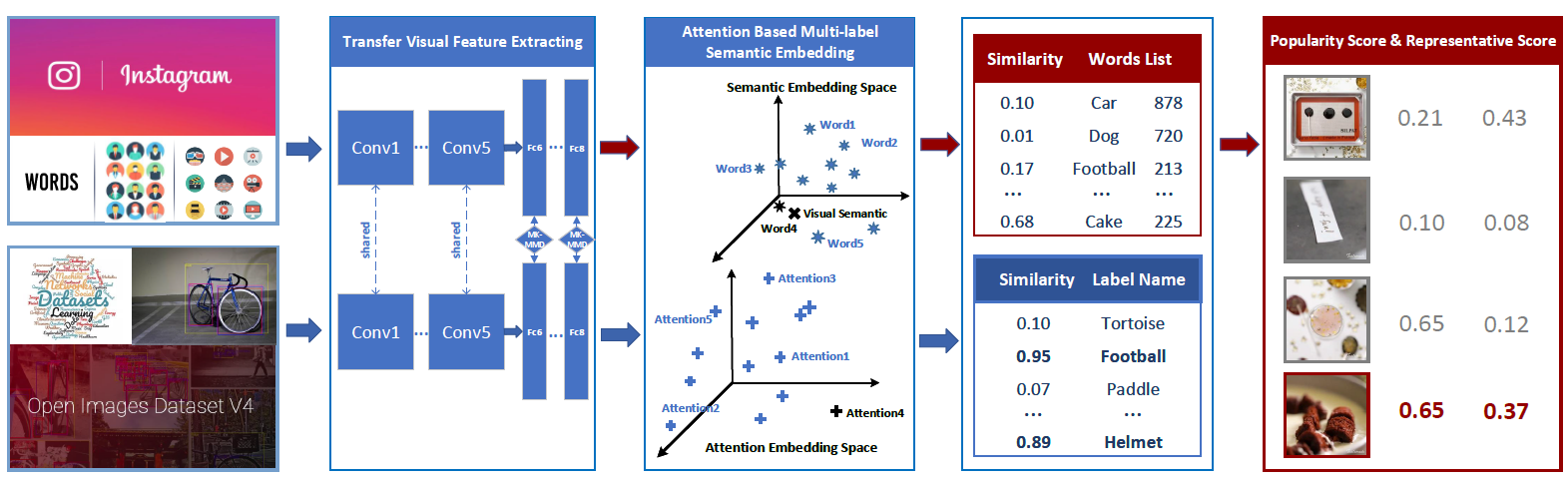}
	  \vspace{-1mm}
	  \caption{Illustration of our framework. The framework is comprised of several components: transfer visual feature extraction, attention based multi-label visual-semantic embedding and popularity calculation. In the training stage, a multi-label image recognition model is trained. And the learned model is transferred to calculate the popularity for each frame by measuring its similarity with the popular topics. We ultimately fuse the popularity and the representativeness scores to yield the attractive thumbnail.}
	\label{fig_2}	  
	\vspace{-5mm}
\end{figure*}

Based on the constructed dataset~\cite{Benchmark}, we devise an Attention based MUlti-label visual-Semantic Embedding model (AMUSE) to measure the similarities between micro-videos and intents. In particular, the attention mechanism is introduced into our proposed model for distinguishing the mixed visual features according to the labels. As shown in Figure 3, leveraging the attention vectors of the labels, some related information is preserved and the unrelated is omitted. The attended feature vectors facilitate the cross-modal similarity computation in the multiple labels case. Nevertheless, due to the same scalability problem, it is also hard to obtain the attention vectors of unseen labels. To solve this problem, we learn an associated attention embedding space and a semantic-attention projection yielding the attention vector for unseen labels, which is inspired by the visual-semantic embedding model.
As to the third component, we sum the similarities weighted by the frequencies of the corresponding words as the popularity score for each frame, and then combine it with the representativeness score to determine the thumbnail. By conducting extensive experiments on our constructed dataset, we demonstrate that our proposed model outperforms several state-of-the-art baselines.

The main contributions of this work are threefold:
\begin{itemize}
    \item We are the first on proposing a scheme to calculate the frame-wise popularity for micro-video thumbnail selection, as well as construct a micro-video dataset associated with the popular topic list. 
    \item To avoid the massive cost of manually annotating the frames, we trained the model on an auxiliary dataset and transferred the learned knowledge to predict the similarity with a deep transfer model.
    \item Considering the unseen words emerging in the popular topic list after the training phase, we devise an attention based multi-label visual-semantic embedding model to calculate the similarities between each frame and all words in the popular topic list. Thereinto, we learn an attention embedding space and a semantic-attention projection to yield the attention vector for any word.

\end{itemize}
\section{Related Work}
Our work is related to a broad spectrum of thumbnail selection and cross-modal similarity computation.
    \begin{figure}
	\centering
	  \includegraphics[width=0.47\textwidth]{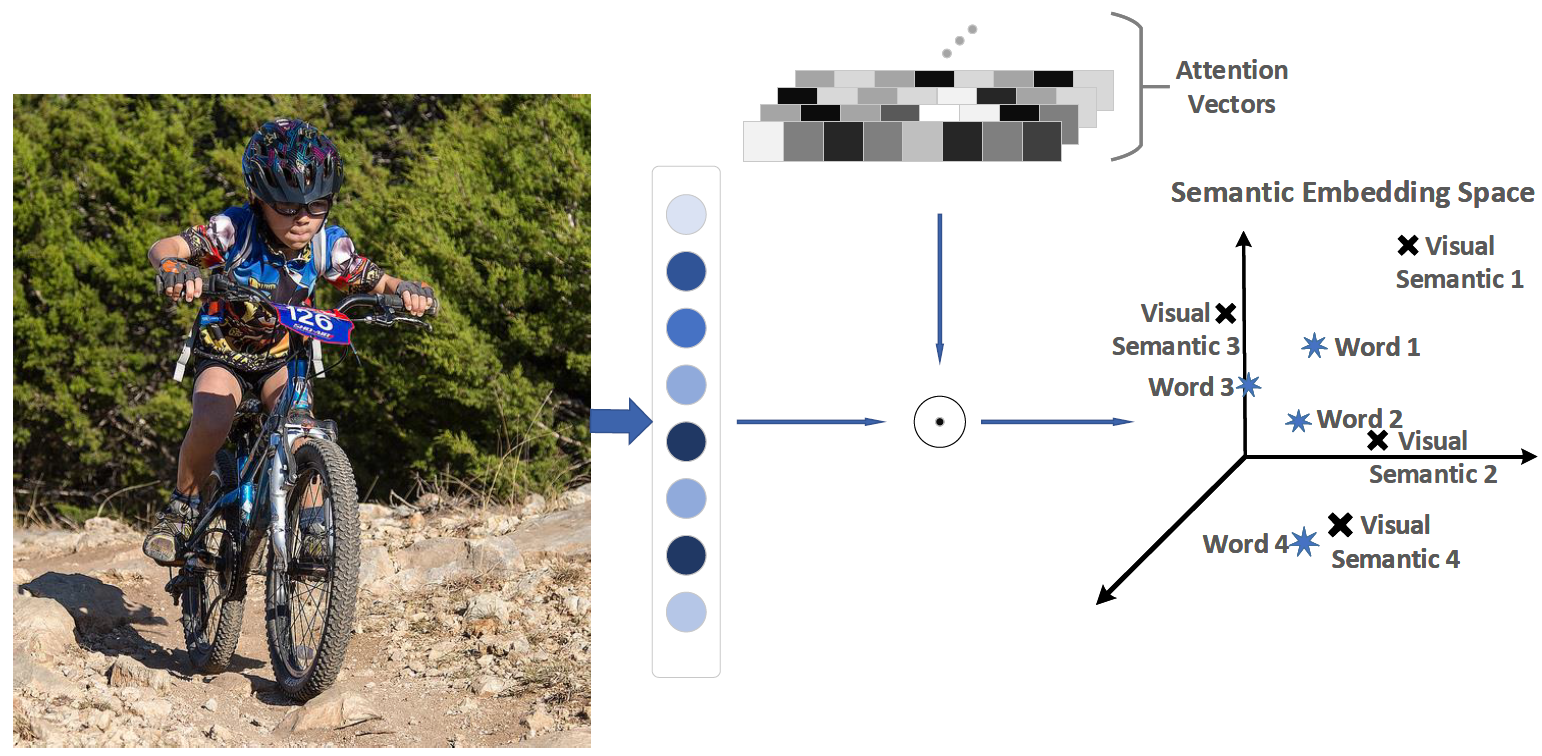}
	  \vspace{-1mm}
	  \caption{Exemplar illustration of the multi-label visual-semantic projection. One feature vector is weighted by multiple attention vectors and projected to several positions in the semantic embedding space.}
	\label{fig_3}	  
	\vspace{-5mm}
\end{figure}

\subsection{Cross Modality Similarity Computation}
Currently, the mainstream in the cross modality similarity computation~\cite{3,4,5,6,7,8,9,10,11,12,13,NMCL} is the common space learning. These methods follow the idea that there is a latent common space where the similarity can be directly measured.

Here, we group these methods into two categories, statistical correlation analysis fashion and deep learning fashion. The former learns the linear projection matrices by optimizing the statistical values. Canonical Correlation Analysis (CCA)~\cite{14} is straightforwardly applied to select the shared latent subspace by maximizing the correlation between the views. Since the subspace is linear, it is impossible to apply CCA to the real-world datasets with non-linearities. To compensate for this problem, Akaho~\cite{15} proposed a kernel variant of CCA, namely KCCA. Rasiwasiz et al.~\cite{16} proposed a model which first applies CCA to yield the common space of visual and
textual information. Besides CCA, Li et al.~\cite{17} devised a Cross-modal Factor Analysis (CFA) to minimize the Frobenius norm between the pairwise data in the common space.

With the success of deep learning in computer vision~\cite{18,19} and natural language processing~\cite{20,21}, some deep learning based approaches have been presented to compute
the cross modality similarity. For instance, Ngiam et al.~\cite{22} employed a deep belief network with the extension restricted Boltzmann machine for the shared feature learning of different modalities and proposed a bimodal deep autoencoder to learn the cross-modal correlations. Yu et al.~\cite{23} used the deep Convolutional Neural Network (CNN) for image embedding, while at the same time kept the transformation at the
textual information. To understand inter-modality semantic correlations, He et al.~\cite{24} designed a deep and bidirectional representation learning model, where images and text are mapped to a common space by two convolution-based networks. Simultaneously, a bidirectional network architecture is devised to capture the property of the bidirectional search. To overcome the category limitations of the conventional
models, Frome et al.~\cite{25} used a pre-trained word2vec model for text embedding, and trained a visual-semantic projection to map the visual information into this semantic space to directly measure the similarity between the information of different modalities.

Although the visual-semantic embedding model has shown impressive results on the computing similarity from the visual information to a single label, it fails to optimize the current model for the multi-label similarity calculation. Therefore, we
proposed a multi-label visual-semantic embedding model to settle this problem.
\section{Methodology}
Our proposed thumbnail selection architecture can be decomposed into three components: 1) characterizing the visual information of each candidate frame via a deep transfer method; 2) projecting the extracted feature vector into the visual-semantic space for popularity calculation; and 3) selecting the thumbnail for each micro-video by considering both the representativeness and popularity. In this section, we detail them in sequence.

\subsection{Notation}

For notations, we use bold capital letters (e.g., $\mathbf{X}$) and bold lowercase letters (e.g., $\mathbf{x}$) to denote matrices and vectors, respectively. In addition, non-bold letters (e.g.,x) are employed to represent scalars. If not clarified, all vectors are in column forms.

Suppose we are given a set of $N$ micro-videos $\mathcal{X}=\{x_i\}_{i=1}^N$ and one list of popular topics $\mathcal{W}=\{w_i\}_{i=1}^M$  consisting of $M$ words. For each micro-video $x \in \mathcal{X}$, several candidate frames are extracted based on the visually high-quality and representativeness. In our model, we define the popularity as the sum of similarity scores computed by the distance from the extracted visual vector $v$ to each word $w$ in a semantic embedding space. The embedded word vector denoted by $s$ in this semantic space is called a prototype. Finally, we combine the representativeness score and the popularity score to select the frame with the highest score as the thumbnail for each micro-video.
    \begin{figure}
	\centering
	  \includegraphics[width=0.4\textwidth]{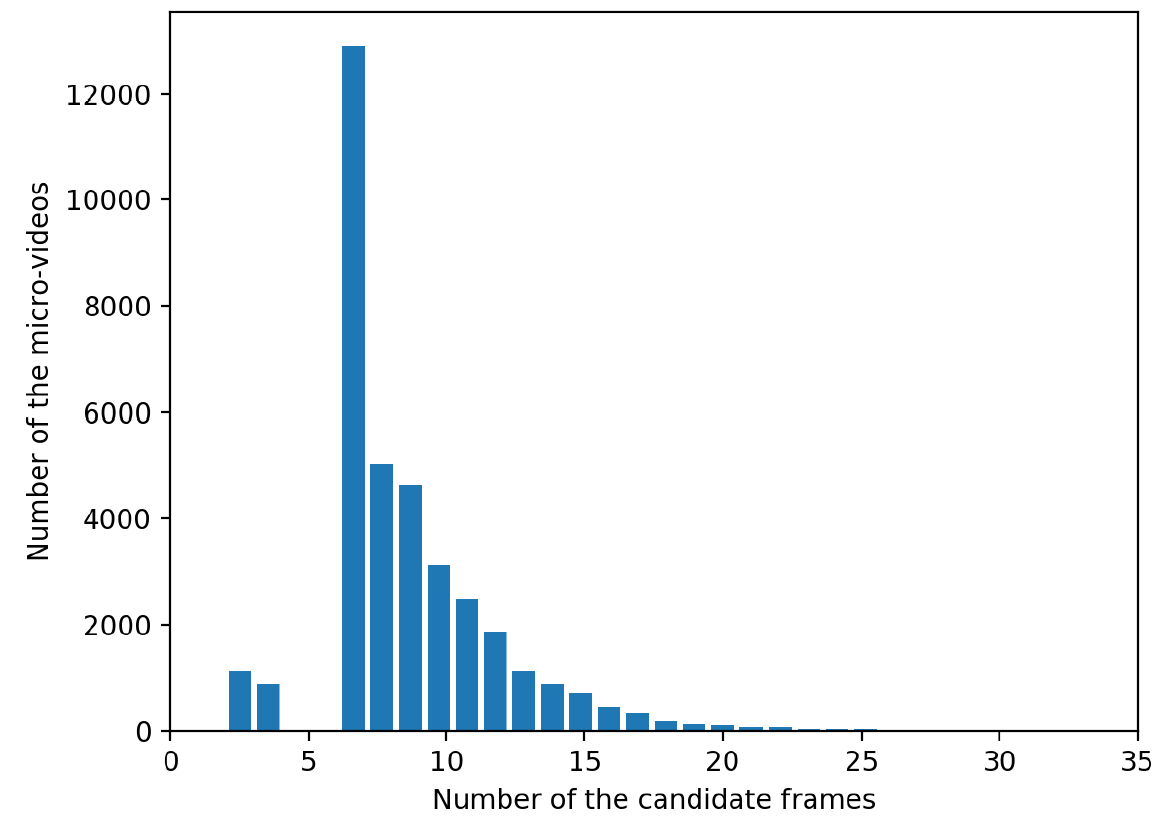}
	  \vspace{-1mm}
	  \caption{The distribution of the extracted candidate frames in our micro-video dataset.}
	\label{fig_5}	  
	\vspace{-5mm}
\end{figure}

\subsection{Deep Transfer Learning}

In our task, manually annotating the large training frames with multiple labels is inapplicable, and directly adapting cross-model similarity calculation model to the extracted frames is impossible. Therefore, a deep transfer network is employed and trained over Open Images Dataset in which the images are annotated by multiple labels. Through the
transfer model, we bridge the gap of the extracted visual features between the auxiliary dataset and the micro-video dataset to better convey the learned knowledge of similarity calculation~\cite{27}.

Long et al.~\cite{28} suggested that the features are extracted from general to specific in the last fully connected layers of CNNs. The transferability decreases with the domain discrepancy when transferring to the higher layers. It means that the fully connected layers are tailored to the specific task; hence the model cannot be directly transferred to the target domain via fine-tuning with limited target supervision. To transfer the knowledge from the source domain to target domain, we train the model on the annotated dataset and reduce the discrepancy between their distributions under the representations of fully connected layers $fc6-fc8$, as shown in Figure 2. This model can be implemented by the multiple kernel maximum mean discrepancies (MK-MMD), as
\begin{equation}
\delta_k^2(p,q)\triangleq\left \|E_p\left[\phi(\mathbf{x}^s) \right]-E_q\left[\phi(\mathbf{x}^t) \right]\right \|_{\mathcal{H}_k}^2,
\end{equation}
where $\mathbf{x}^s$ and $\mathbf{x}^t$ denote the inputs of the auxiliary dataset and the micro-video dataset, respectively. The $\mathcal{H}$ denotes the reproducing kernel Hilbert space endowed with a character kernel $k$. And the mean of distribution $p$ in $\mathcal{H}_k$ is a unique element ${\mu_k}(p)$ such that $E_p\left[f(x)\right]=\left \langle f(x),{\mu_k}(p)\right \rangle$ for all $f\in\mathcal{H}_k$. In addition, the characteristic kernel associated with the feature map $\phi$, $k(\mathbf{x}^s,\mathbf{x}^t)=\left \langle \phi(\mathbf{x}^s),\phi(\mathbf{x}^t)\right \rangle$is defined as the convex combination of $m$ kernels $k_u$,
\begin{equation}
\mathcal{K}\triangleq\{k=\sum_{u=1}^m \beta_u k_u;{\sum_{u=1}^m \beta_u}=1,\beta_u\ge 0,\forall u\},
\end{equation}
where the constraints on coefficients $\beta_u$ are imposed to guarantee that the multi-kernel $k$ is characteristic. Ultimately, benefiting from this deep transfer model, we obtain a $d$-dimensional feature vector for each image/frame.
    \begin{figure}
	\centering
	  \includegraphics[width=0.4\textwidth]{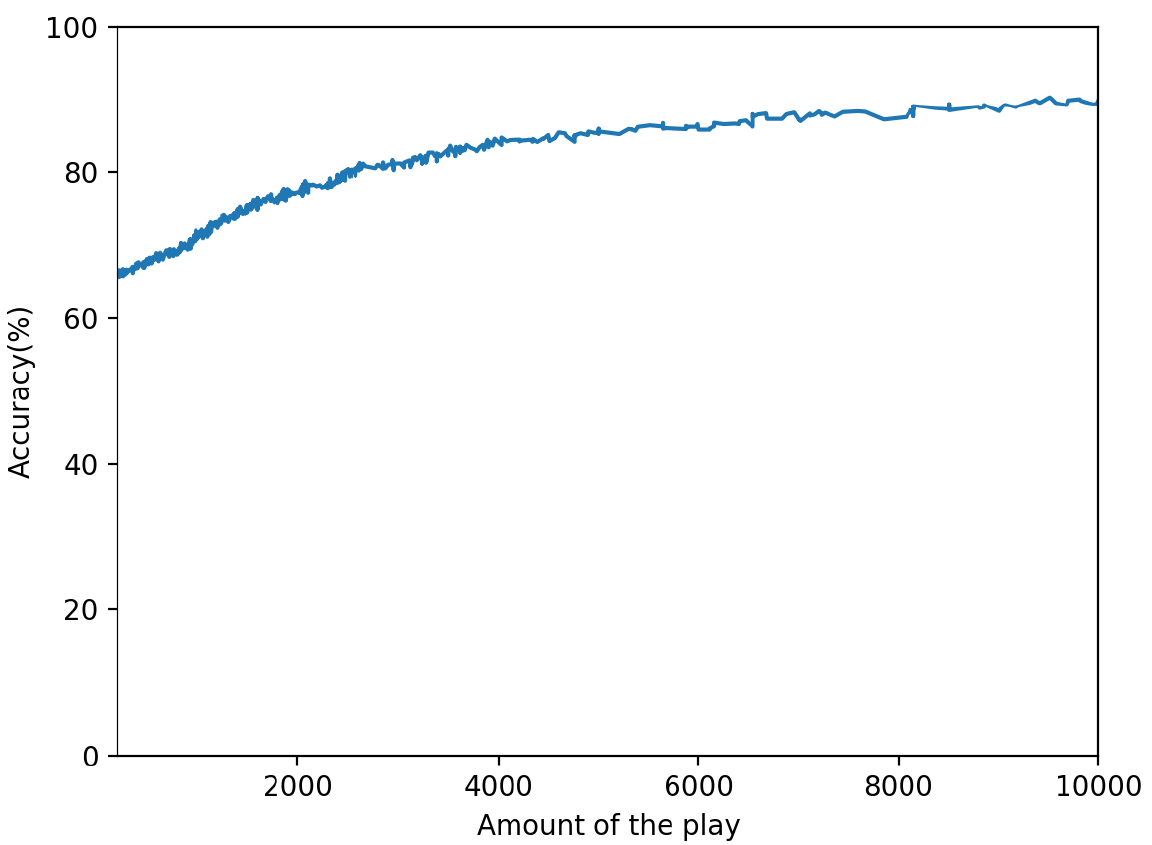}
	  \vspace{-1mm}
	  \caption{Manually labeling accuracy statistics over 1,000 micro-videos.}
	\label{fig_6}	  
	\vspace{-5mm}
\end{figure}

\subsection{Multi-label Visual-Semantic Embedding Model}

Our proposed model aims to measure the popularity of the frames extracted from the micro-video. Here, the popularity of one frame is defined as the sum of the similarity scores between the visual information and each word in the popular topic list
~\cite{Jiang}.

To learn and transfer the knowledge of cross-modal similarity measurement, we train a multi-label recognition model related to our similarity calculation task over the auxiliary dataset. Nevertheless, the standard recognition models lack the scalability over the number of class labels, and need to be retrained when any new label emerges. Considering the variability of the popular topics, it is hard to directly apply these recognition models to our task. Hence, we apply a visual-semantic embedding based model to learn a continuous semantic space which captures the semantic relationship among labels and explicitly learns the mapping function from visual features to the semantic space.

Although this model resolves the drawback of the standard recognition model in scalability, it is nontrivial to extend a single-label visual-semantic model to a multi-label one. Towards this end, we devise a multi-label visual-semantic embedding model trained by the auxiliary dataset consisting of pairs between the image and multiple labels. Moreover, the attention mechanism is adopted to reweight each visual feature vector, since the relativeness between one feature and different labels may vary. Regarding the different labels, corresponding attention vectors are used to enhance the related feature and filter out the irrelevant ones. With the help of the attention mechanism, the information related to the label is preserved and the feature vectors can be mapped to different coordinates in the semantic space for different labels.

\begin{table}
  \centering
  \caption{List of the representative popular topics and their frequencies in our dataset.)}
    %\vspace{0.1mm}
  \label{table_1}
  \setlength{\tabcolsep}{2.0mm}
  \begin{tabular}{|c|c|c|c|}
    \hline
    \textbf{Word}&\textbf{Times}&\textbf{Word}&\textbf{Times}\\
    \hline
    Girl&29211&Beach&2388\\
    \hline
    Baby&15873&Cake&2225\\%$48,888$ micro-videos, $2,303$ users, and $12,194$
    \hline
    Boy&1048&Guitar&2122\\
    \hline
    Car&8783&Football&2013\\
    \hline
    Dog&7230&Child&1982\\
    \hline
    Gym&7001&Chocolate&1790\\
    \hline
    Family&6154&Bike&1678\\
    \hline
    Smoke&4873&Bag&1524\\
    \hline
  \end{tabular}
  \vspace{-2mm}
\end{table}

Similar to the disadvantage of the traditional recognition model~\cite{Yu}, the attention vector is also difficult to scale, especially when the words in the popular topic list are alterable. Inspired by the visual-semantic embedding model, we introduce the attention embedding space and learn a semantic-attention projection. With this projection, the attention vector of any prototype can be obtained and used to weight the visual feature
vectors. In what follows, we elaborate the critical ingredients of the proposed model.

1) Semantic Embedding Space Constructing: To learn the semantically-meaningful representations for each term from the unannotated text, Mikolov et al.~\cite{29} introduced a skipgram text modeling architecture. This model represents the word as a fixed length embedding vector by predicting its adjacent terms in the document. Since synonyms appear in similar contexts, this simple objective function drives the model to learn similar embedding vectors for semantically related words.

To construct the semantic space, we utilize the Global Vectors for Word Representation (GloVe) model~\cite{30} trained with the unannotated text data from the Instagram. This model has been demonstrated to map the words to the semantically-meaningful embedding features and learns the similar embedding points for semantically related words since synonyms have similar semantic contexts.

2) Attention Embedding Space Constructing: Attention mechanism has been proven to be a practical approach for embedding categorical inference applications within a deep neural network~\cite{31,32,33}. It reasonably assumes that human recognition tends to focus on the parts that are needed instead of the whole perception space. With this assumption, this approach alleviates the bottleneck of compressing a source into a fixed-dimensional vector by equipping a model with variable-length attention memory~\cite{34}.

Attention network is defined to access a memory matrix which stores the useful information to solve the task at hand. In the current recognition task, there is an attention matrix denoted as $A\in\mathbb{R}^{C\times d}$, where $C$ and $d$ represent the number of the categories and the dimension of the attention vector, respectively. The matrix is constructed to store the attention score of each feature for the certain categories and used to weight the feature vector to yield an attended feature vector, as
   \begin{figure*}
	\centering
	  \includegraphics[width=0.8\textwidth]{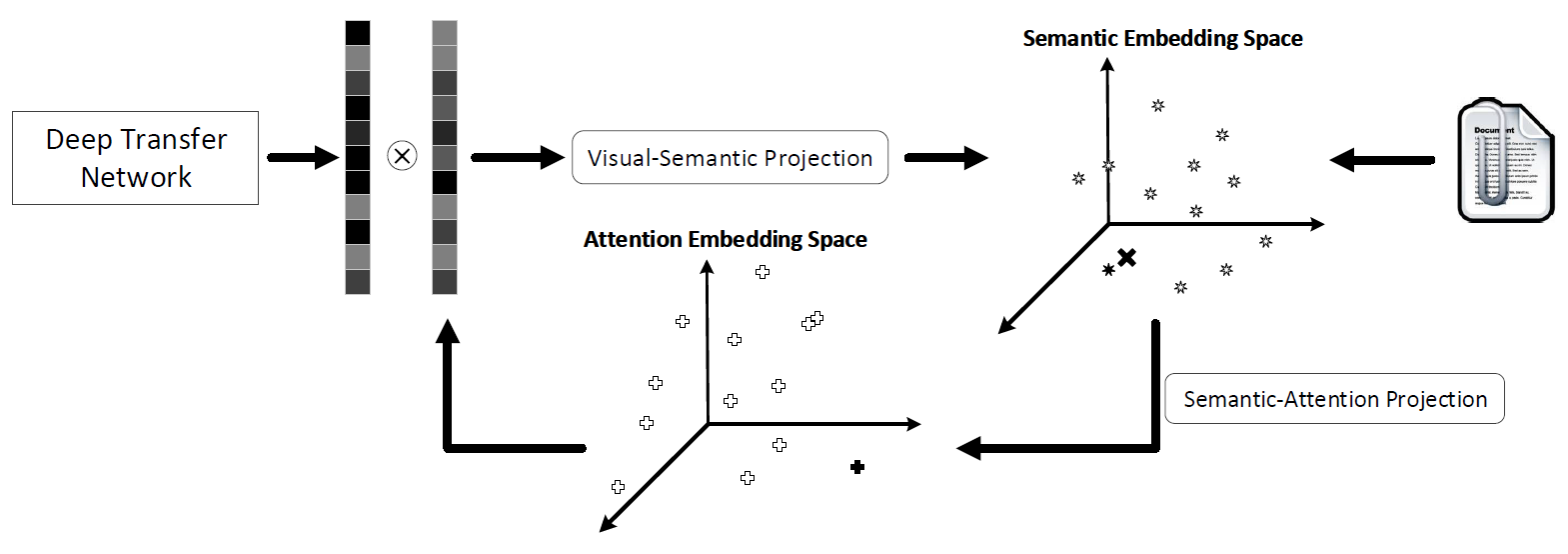}
	  \vspace{-1mm}
	  \caption{Illustration of the multi-label visual-semantic embedding model. To measure the distance between the visual feature vector with various words, attention embedding space and semantic embedding space are constructed. We map the attended feature vectors towards corresponding prototypes. Finally, the distance between each attended vector and prototype is considered as their similarity.}
	\label{fig_7}	  
	\vspace{-5mm}
\end{figure*}

\begin{equation}
\widehat{v_i}=A_i \otimes v,
\end{equation}
where $v_i$ and $v$ denote the attended feature vector of $i$-th category and the visual feature vector, respectively; $A_i$ is the $i$-th row of the attention matrix and each element means the relevance between the corresponding feature and the $i$-th category.

Here, this matrix is defined before the training process where the size is fixed. It should be retrained for the new category. Nevertheless, in our task, the words in the popular topic list are dynamical, and for some topic words, it is hard to collect enough micro-videos for the training. This traditional attention network is not applicable in this case. Therefore, inspired by the visual-semantic embedding model, we introduce attention embedding space and learn a semantic-attention projection to map the semantic vector to the attention embedding space, as shown in Figure 7, formulated as
\begin{equation}
a_w=\psi(W_a \cdot s_w)+\gamma\left \| W_a \right \|_F^2, 
\end{equation}
where $a_w$, $s_w$, $W_a$ and  $\psi(\cdot)$ denote the attention embedding vector of word $w$, the prototype of word $w$, the weight matrix of trainable parameters and the semantic-attention mapping function, respectively.

3) Attention based Visual-Semantic Mapping: As mentioned above, to measure the popularity of each candidate frame, we calculate the sum of similarity scores between the frame and all words in the popular topic list. For this, each word is represented by a prototype and the frames are projected into the same semantic embedding space. To calculate the distance between the visual vector and each prototype, we have to preserve the features related to the word. Therefore, the attention vectors are leveraged to enhance the associated features and omit the unrelated ones, as
\begin{equation}
\widehat{v_w}=a_w \otimes v,
\end{equation}
In this formulation, $a_w$ denotes the attention vector yielded by Equation 4 of input word $w$ and $\widehat{v_w}$ is the attended feature vector towards the word $w$. As shown in Figure 7, we map all attended feature vectors into the semantic space. To train this multi-label visual-semantic embedding model, we minimize a function combining the dot-product similarity and the hinge rank loss, as
\begin{equation}
\begin{split}
  \mathcal{J}(v,\mathcal{W})=&\sum_{w\in\mathcal{W}} \sum_{j\neq w} max\left[0,\alpha -{s_w}{\mathbf{W}_v}\hat{v}_w+{s_j}{\mathbf{W}_v}\hat{v}_w\right]\\
  &+\eta\left \| W_v \right \|_F^2,  
\end{split}
\end{equation}
where $w$ is the word from the set $\mathcal{W}$, $\alpha$ and $\mathbf{W}_v$ denote the margin and the matrix of trainable parameters in the linear transformation layer, respectively.

At the whole training stage, the final objective function consists of the multi-label recognition empirical risk and MK-MMD distance, as
\begin{equation}
\mathcal{L}=\mathcal{J}(v,\mathcal{W})+\mu \sum_{l=l_1}^{l_2}\delta_k^2{(p,q)},
\end{equation}
where $\mu$ denotes a penalty parameter, $l_1$ and $l_2$ are layer indices between which the MK-MMD distance is calculated.

\begin{table}
  \centering
  \caption{Performance comparison between our model and
the baselines.}
    %\vspace{0.1mm}
  \label{table_2}
  \setlength{\tabcolsep}{2.0mm}
  \begin{tabular}{|c|c|}
    \hline
    \textbf{Method}&\textbf{Accuracy}\\
    \hline
    \textbf{Random}&$14.09\pm 2.81\%$\\
    \hline
    \textbf{ASBTV}\cite{26}&$20.92\pm 0.31\%$\\
    \hline
    \textbf{DeViSE}\cite{25}&$24.09\pm 0.35\%$\\
    \hline
    \textbf{MTL-VSEM}\cite{9}&$27.01\pm 0.57\%$\\
    \hline
    \textbf{C2AE}\cite{35}&$26.16\pm 0.53\%$\\
    \hline
   \textbf{AMUSE}&\bm{$30.10\pm 0.29\%$}\\
    \hline
  \end{tabular}
  \vspace{-2mm}
\end{table}

\subsection{Thumbnail Selection}

In the inference phase, we utilize the trained model to characterize the visual information of each frame and calculate their popularity scores. Hence, after obtaining the popular topic list, we map the words in the list into the semantic space and receive the prototypes. And then, each prototype’s attention vector is yielded via the trained semantic-attention projection, and used to weight the visual features for attended feature vectors. Sequentially, they are mapped to various coordinates in the semantic embedding space towards their prototypes. The sum of frequency-based weighted similarities is treated as the popularity score of each frame, formulated as
\begin{equation}
P(v,\mathcal{M})=\sum_{m\in\mathcal{M}} {T_m}\cdot {s_m}{\mathbf{W}_v}\hat{v}_w,
\end{equation}
where $T_m$ denotes the frequency of the $m$-th word in the popular topic list. Finally, we fuse the visual representativeness score and popularity score of each candidate frame, as
\begin{equation}
S(x_i,\mathcal{M})=P(v_i,\mathcal{M})+\lambda \cdot R(x_i),
\end{equation}
where $R(\cdot)$ and $S(\cdot)$ denote the representativeness score of the micro-video $x_i$ and the sum of representativeness and popularity, respectively. Accordingly, we choose the one with the highest score as the thumbnail of the micro-video, as shown in Figure 2.
\section{EXPERIMENTS}
In this part, we carried out extensive experiments to thoroughly validate our proposed model and its components on the constructed micro-video dataset.

\subsection{Parameter Settings}
For the deep transfer model, we employed the AlexNet-based deep transfer model introduced in~\cite{36}. Notably, the model extends the AlexNet architecture comprised of five convolutional layers (conv1 - conv5) and three fully connected layers (fc6 - fc8). The pre-trained AlexNet is adapted, whose conv1-conv3 layers have been frozen and the conv4 - conv5 layers would be fine-tuned at the training stage. Besides, we applied Xavier approach to initializing the model parameters, proven as an excellent initialization method for the neural networks. 
Following the prior work~\cite{HUIGN}, the mini-batch size and learning rate are respectively searched in \{128, 256, 512\} and \{0.0001, 0.0005, 0.001, 0.005, 0.01\}. The optimizer is set as Adam. Moreover, we empirically set the size of each hidden layer as 256 and the activation function as ReLU. Without special mention, except the deep transfer network, other models employ one hidden layer and one prediction layer. For two embedding space construction, we explored the dimensionality of them in \{50-500\} and \{1024, 2048\}, respectively. For a fair comparison, we initialized other competitors with the analogous procedure. We showed the average result over five-round predictions in the testing set.
  \begin{figure*}
	\centering
	  \includegraphics[width=1\textwidth]{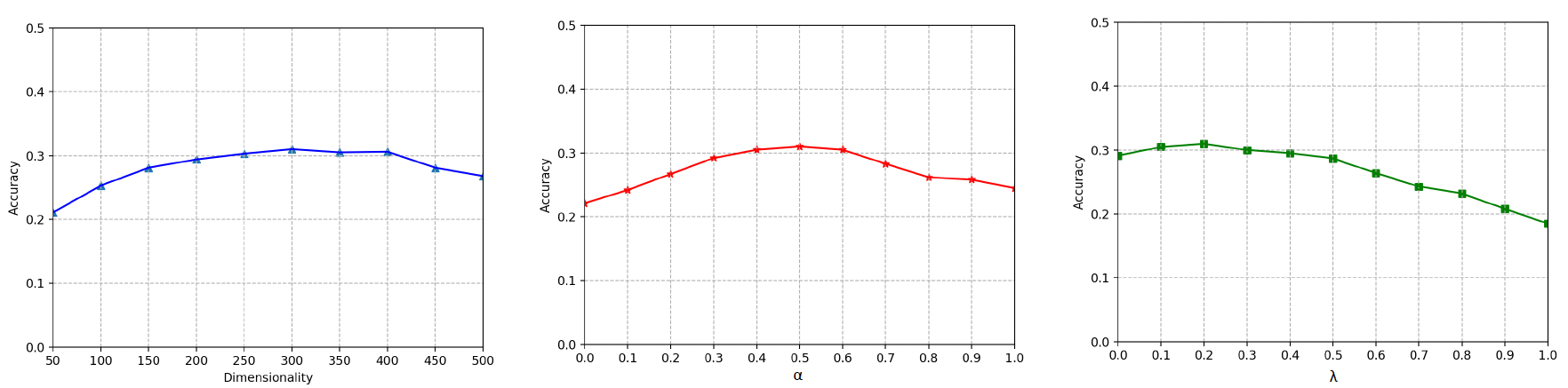}
	  \vspace{-1mm}
	  \caption{Performance of our proposed model by varying parameters.}
	\label{fig_8}	  
	\vspace{-5mm}
\end{figure*}

\begin{figure}
	\centering
	  \includegraphics[width=0.4\textwidth]{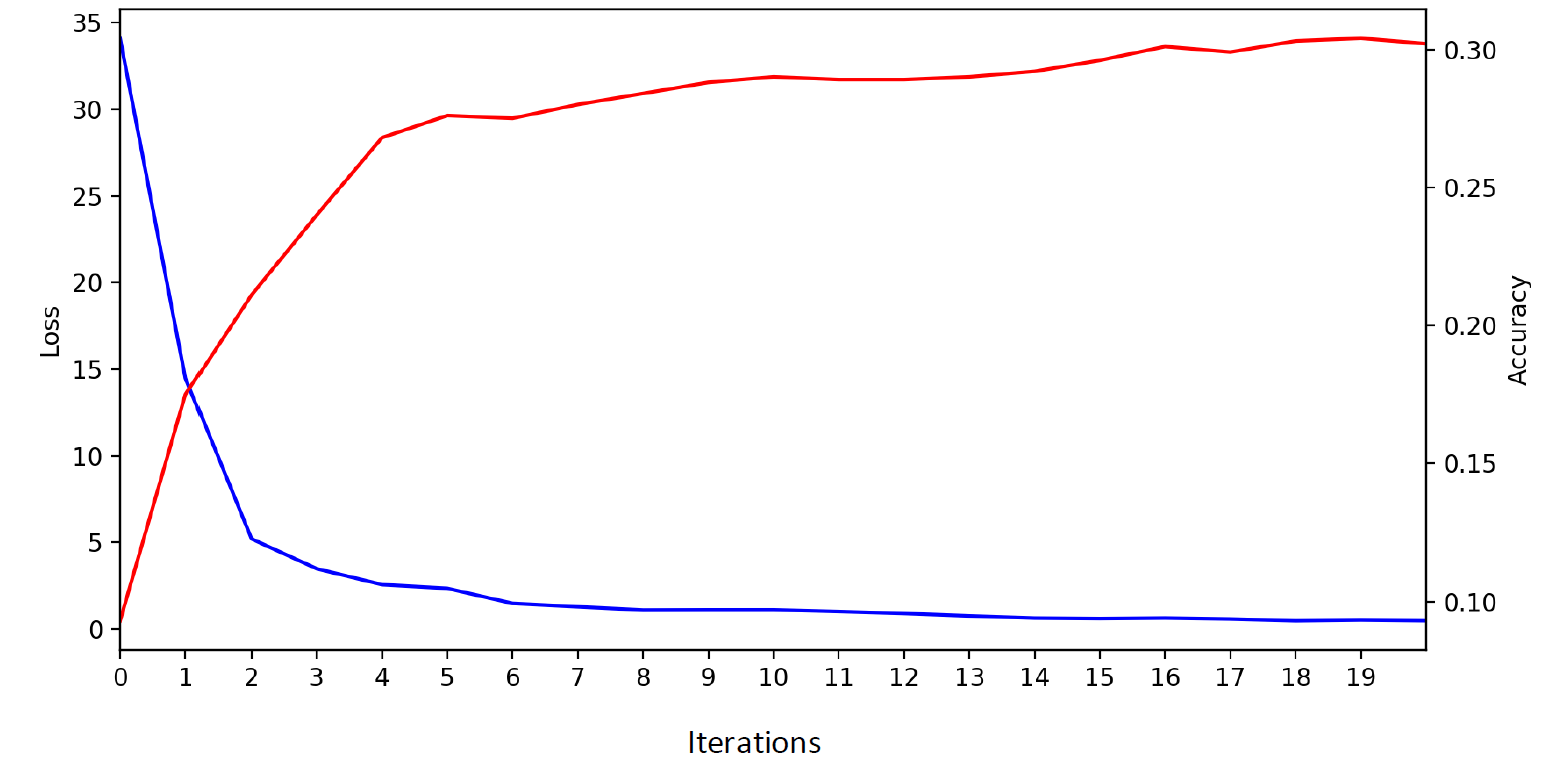}
	  \vspace{-1mm}
	  \caption{Performance of our proposed model with the iteration times.}
	\label{fig_9}	  
	\vspace{-5mm}
\end{figure}

\subsection{Baselines}

To evaluate our proposed model for the thumbnail selection, we compared it to the following six methods on the labeled micro-video dataset, including Random, ASBTV~\cite{26},  DeViSE~\cite{25}, MTL-VSEM~\cite{9}, and C2AE~\cite{35}.
% \begin{itemize}
%     \item Random. The method randomly selects one frame from the candidate frame set as the thumbnail.
%     \item ASBTV~\cite{26}. The authors proposed an automatic thumbnail selection system to exploit the representation and the visual attractiveness of frames. This system measures the
% attractiveness by analyzing the visual quality and aesthetic metrics, as well as the representativeness achieved by their proposed clustering algorithm.
%     \item DeViSE~\cite{25}. The model is the first to construct a semantic embedding space through the natural language model and map the visual information into this space to resolve the scalability problem of standard recognition models. We leveraged this model to select an attractive thumbnail for each micro-video via popularity computation.
%     \item MTL-VSEM~\cite{9}. The authors developed a multi-task deep visual-semantic embedding model to automatically select the query-dependent video thumbnails according to both the visual and side-information. We retrained this model to compute the relevance between the frame and popular topics, instead of the visual information and queries.
%     \item C2AE~\cite{35}. The authors proposed a novel deep neural network (DNN) based end-to-end model for solving multilabel classification task. This model integrates the DNN architectures of the canonical correlation analysis and autoencoder, allowing the learning and prediction with the ability to exploit label dependency.
% \end{itemize}
\subsection{Performance Comparison}
The comparative results are shown in Table II. From this table, we have the following observations:
\begin{itemize}
    \item  In terms of accuracy, the random method performs the worst, indicating the significance of the video content.
    \item  The side-information based models (e.g., MTL-VSEM and C2AE) outperform ASBTV which only considers the visual quality and representativeness. This result demonstrates that the side-information can improve the micro-video understanding.
    \item  When performing the thumbnail selection task, MTLVSEM is superior to DeViSE. It is reasonable since MTL-VSEM employs the multi-task learning which refers to the joint training of multiple tasks, while enforces a common intermediate parameterization or representation to improve each task’s performance.
    \begin{table}
  \centering
  \caption{Performance comparison between our model and
several variants.}
    %\vspace{0.1mm}
  \label{table_3}
  \setlength{\tabcolsep}{2.0mm}
  \begin{tabular}{|c|c|}
    \hline
    \textbf{Method}&\textbf{Accuracy}\\
    \hline
    \textbf{Variant-I}&$25.34\pm 0.09\%$\\
    \hline
    \textbf{Variant-II-seen}&$28.67\pm 0.15\%$\\
    \hline
    \textbf{Variant-II-unseen}&$20.45\pm 1.57\%$\\
    \hline
    \textbf{Variant-III}&$26.24\pm 0.25\%$\\
    \hline
    \textbf{Variant-IV}&$28.30\pm 0.11\%$\\
    \hline
   \textbf{AMUSE}&\bm{$30.10\pm 0.29\%$}\\
    \hline
  \end{tabular}
  \vspace{-2mm}
\end{table}
    \item  C2AE surpasses the visual-semantic embedding based approach DeViSE. This verifies that the multi-label task benefits our thumbnail selection, since the popularity calculation should consider the similarities between the pairs of each frame and all words in the list of popular topics. And the standard visual-semantic embedding model causes the more significant error on the distances calculation to multiple prototypes.
    \item  It is observed that MTL-VSEM outperforms C2AE. The discrepancy of the training data and testing data is the primary cause, while C2AE ignores the gap between them, and MTL-VSEM integrates the multi-task learning to decrease this discrepancy. In addition, C2AE cannot be extended to the unseen prototypes, leading to the external error.
    \item  Our proposed model performs the best. By introducing the attention based multi-label part, our model achieves better expressiveness on calculating the similarity between the visual feature vector to various prototypes. Besides, we employed the deep transfer learning to reduce the gap between the source domain and target domain and transfer the knowledge from the former to the latter.
\end{itemize}
\subsection{Study of AMUSE Model}

1) Study of Components: In this section, we studied the effectiveness of each component of our proposed model. We listed the following several variants to compare with the proposed model.
\begin{itemize}
    \item Variant-I. In this model, we leveraged a pre-trained recognition model to replace the deep transfer learning components. The model is fine-tuned on the Open Images Dataset for the multi-label learning and embeds the visual information into the semantic embedding space. During the multi-label training, the attention embedding space and the semantic-attention projection are learned. This variant is designed to investigate the effectiveness of the deep transfernlearning model for our task.

    \item Variant-II. It focuses on the effectiveness of the observed words and unobserved words. For this, instead of modifying the proposed model, we divided the popular topics into two groups, including the seen words and the unseen words. We respectively harnessed these two groups to select the thumbnail for the micro-videos and compared them to the result achieved by all words.

    \item Variant-III. This variant discards the attention mechanism and directly projects the visual information into the semantic space. Hence, the single coordinate should be compared to all prototypes. This experiment is used to verify that the single embedded visual vector degrades the popularity measurement.

    \item Variant-IV. In this variant model, we removed the visual-semantic embedding and mapped the multi-modal information into a common latent space for distance computation. With the visual-semantic space removed, it is hard to measure the similarity between the visual information and unseen words. Besides, the attention vectors of the unseen words could not be achieved. For fairness, we experimented only with the seen words, like variant-II, and compared our proposed model with this variant.

\end{itemize}
In Table III, we have the following observations:
\begin{itemize}
    \item Regarding the accuracy, our proposed approach outperforms Variant-I. Because we exploited the deep transfer network to bridge the gap between the source domain data and target domain data, where the discrepancy between them is reduced. The feature extracted from the target data is appropriate for the subsequent training parts.

    \item From the comparison between Variant-II-seen and Variant-II-unseen, we found that the accuracy of applying the seen words is much better because of the projection domain shift. However, the knowledge of visual-semantic projection learned from the seen word-image pairs can be applied to map the unseen words into semantic embedding space.

    \item Variant-III is devised to evaluate the influence of the attention mechanism for our proposed model. From the result, we empirically showed the effectiveness of the attention mechanism, which not only implements the multi-label visual-semantic embedding but also identifies the importance of visual features.

    \item The last variant, Variant-IV, is introduced to study the effect of the semantic-visual information. We tried to replace the visual-semantic projection by a latent common space associated with corresponding projection. In this common space, the similarity between the visual information and textual information can be computed directly, whereas the correlations between the prototypes are neglected. Although the latent common space based model can be trained to give the comparable performance, the visual-semantic embedding model can leverage the unseen words to improve the accuracy.

    \item Our proposed method significantly outperforms its all variants, justifying that our approach is rational and effective. Different from several variants, the original one considers the gap between the source domain and target domain and leverages the multi-label visual-semantic embedding model for the similarity computation.
     \begin{figure*}
	\centering
	  \includegraphics[width=1\textwidth]{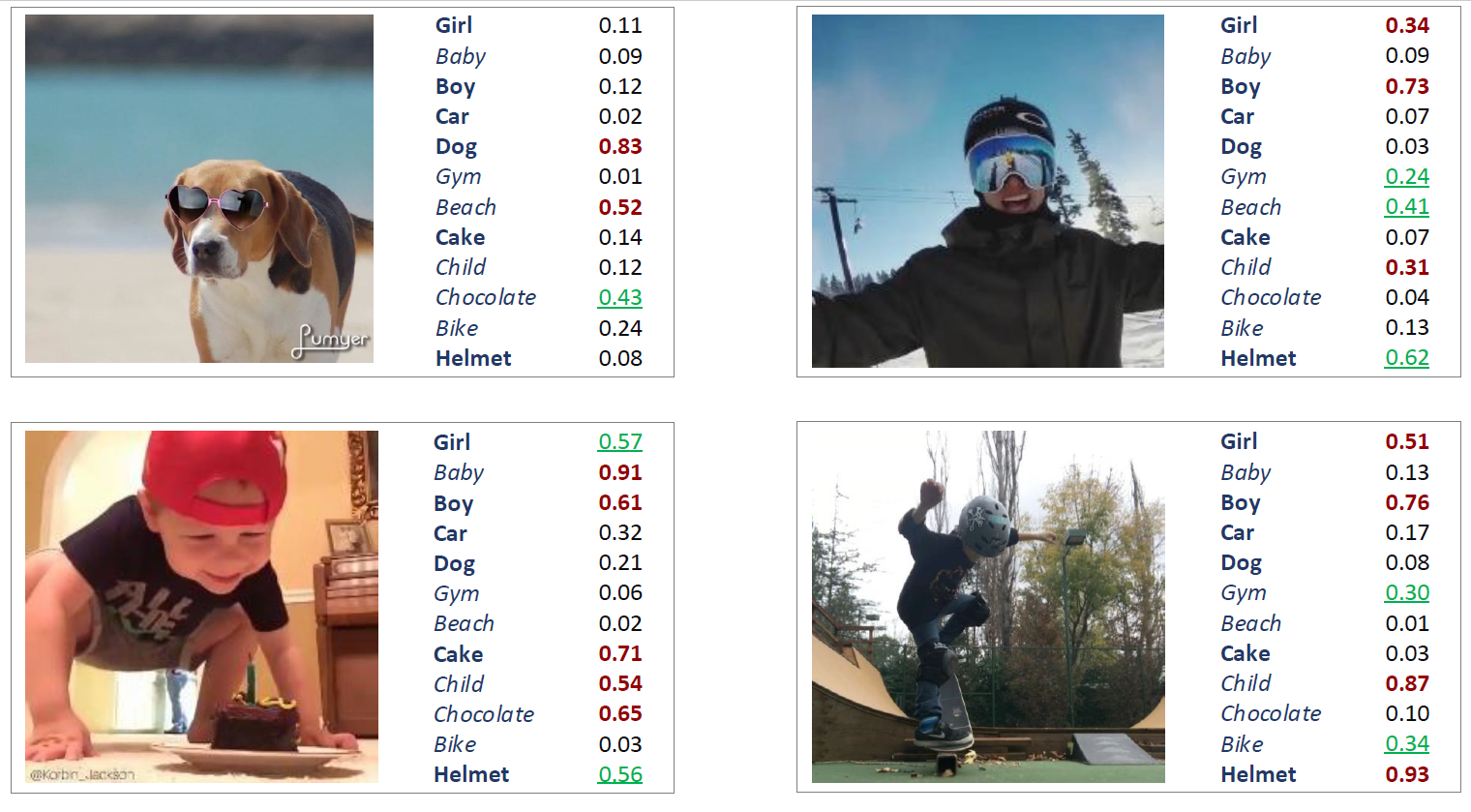}
	  \vspace{-1mm}
	  \caption{Visualization of the similarity scores between the frame and popular topics.}
	\label{fig_10}	  
	\vspace{-5mm}
\end{figure*}

\end{itemize}
2) Parameter Tuning and Sensitivity: We have three key hyper-parameters to study, namely the dimensionality d of the semantic embedding space, margin value in Equation 6 and Equation 9. The optimal values of these parameters are tuned with 5-fold cross-validation. In particular, for each of the 5- fold, we chose the optimal parameters via the grid search with small but adaptive step size. Our parameters are searched in the range of [50, 300], [0, 1] and [0, 1], respectively. The parameters corresponding to the best accuracy are used to report the final results. For other competitors, the procedures to tune the parameters are analogous to ensure fairness.

 Take the parameter tuning in one of the 5-fold as an example. We observed that our model reaches the optimal performance when $d = 300$, $\alpha = 0.5$ and $\lambda = 0.2$. In particular, the parameter $\lambda$ largely affects the performance of our proposed model, and the accuracy is largely affected by the representativeness. It decreases with $\lambda$ increasing, reaches the lowest at $\lambda = 1$ and the best when $\lambda = 0.2$. It is verified that the thumbnail selection should combine the representativeness and the popularity.

We then investigated the sensitivity of our model to these parameters by varying one and fixing the others. Figure 8 illustrates the performance of our model with respect to $d$, $\alpha$ and $\lambda$. We can see that: 1) When fixing $\lambda$, $d$ and tuning $\alpha$ or fixing $\lambda$, $\alpha$ and tuning $d$, the accuracy changes in a relatively small range. The slight change demonstrates that our model is non-sensitive to parameters $d$ and $\alpha$. And 2) Varying $\lambda$ from 0 to 1, the accuracy reaches the peak and then decreases significantly, which means the model is sensitive to the $\lambda$. At last, we recorded the accuracy along with the iteration time using the optimal parameter settings. Figure 9 shows the convergence process concerning the number of iterations. From this figure, it can be seen that our algorithm converges very fast.

\section{Conclusion and Future Work}
In this paper, we presented an automatic thumbnail selection approach towards the micro-video. It combines the visual quality, representativeness, and popularity to select the attractive thumbnail for each micro-video. In this model, we harnessed the aesthetic metrics and the clustering algorithm to select some high-quality and representative frames as the candidates. Besides, a novel multi-label visual-semantic embedding model is proposed to calculate the popularity of these candidate frames. In our model, the popularity is defined as the sum of similarity scores to represent the relevance between each frame and all words in the popular topic list. Towards this end, we constructed a semantic embedding space where the similarity could be calculated directly. To compare the visual-semantic vector with different prototypes, we introduced an attention embedding space associated with the semantic-attention projection. Then, the visual features characterized by a deep transfer model are weighted according to the target words and embedded to various coordinates in the semantic space. Ultimately, the popularity score is obtained and combined with the representativeness score to determine the thumbnail for the micro-video. Extensive comparative evaluations validate the advantages of our proposed model over several state-of-the-art models.

In the future, we expect to capture more elements affecting the thumbnail selection, such as the publisher’s information. Also, we plan to investigate a personalized thumbnail recommendation model which can yield the different thumbnails for the users.
\bibliography{BIB/IEEEabrv, reference}

\end{document}